\documentclass[aoas]{imsart}

\RequirePackage{amsthm,amsmath,amsfonts,amssymb}
\RequirePackage[authoryear]{natbib}
 \usepackage{xr-hyper} 

\RequirePackage[colorlinks,citecolor=blue,urlcolor=blue]{hyperref}
\RequirePackage{graphicx}
\usepackage{booktabs}
\usepackage{array}
\usepackage{algorithmic}
\usepackage[ruled]{algorithm2e}
\usepackage[flushleft]{threeparttable}
\usepackage{mathrsfs} 
\startlocaldefs

\theoremstyle{plain}

\newtheorem{theorem}{Theorem}[section]

\theoremstyle{remark}

\newtheorem*{example}{Example}

 \newtheorem{remark}{Remark}[section]

 \newcommand{\qcr}{\fontfamily{qcr}\selectfont}

 \newcommand{\mbf}{\mathbf}

 \newcommand{\E}{\mbox{E}}
 \newcommand{\V}{\mbox{Var}}

 \newcommand{\ba}{\begin{array}}
 	\newcommand{\ea}{\end{array}}

 \newcommand{\tr}{\text{tr}}

\endlocaldefs

\begin{document}

\begin{frontmatter}
\title{Online Sequential Leveraging Sampling Method for Streaming Autoregressive Time Series with Application to Seismic Data}
\runtitle{Sequential Leveraging Sampling}

\begin{aug}
\author[A]{\fnms{Rui} \snm{Xie}\ead[label=e1]{rui.xie@ucf.edu}},
\author[B]{\fnms{T. N.} \snm{Sriram}\ead[label=e2,mark]{tn@uga.edu}}
\author[C]{\fnms{Wei Biao} \snm{Wu}\ead[label=e3]{wbwu@galton.uchicago.edu}}
\and
\author[B]{\fnms{Ping} \snm{Ma}\ead[label=e4,mark]{pingma@uga.edu}}
\address[A]{Department of Statistics and Data Science, and College of Nursing, 
University of Central Florida,
\printead{e1}}

\address[B]{Department of Statistics, University of Georgia,
\printead{e2,e4}}

\address[C]{Department of Statistics,
University of Chicago,
\printead{e3}}

\end{aug}

\begin{abstract}
Seismic data contain complex temporal information that arrives at high speed and has a large, even potentially unbounded volume.
The explosion of temporally correlated streaming data from advanced seismic sensors poses analytical challenges due to its sheer volume and real-time nature.
Sampling, or data reduction, is a natural yet powerful tool for handling large streaming data while balancing estimation accuracy and computational cost. Currently, data reduction methods and their statistical properties for streaming data, especially streaming autoregressive time series, are not well-studied in the literature. In this article, we propose an online leverage-based sequential data reduction algorithm for streaming autoregressive time series with application to seismic data. 
The proposed \textit{Sequential Leveraging Sampling} (SLS) method selects only one consecutively recorded block from the data stream for inference. While the starting point of the SLS block is chosen using a random mechanism based on streaming leverage scores of data, the block size is determined by a sequential stopping rule. 
The SLS block offers efficient sample usage, as evidenced by our results confirming asymptotic normality for the normalized least squares estimator in both linear and nonlinear autoregressive settings.
The SLS method is applied to two seismic datasets: the 2023 Turkey-Syria earthquake doublet data on the macroseismic scale and the Oklahoma seismic data on the microseismic scale.
 We demonstrate the ability of the SLS method to efficiently identify seismic events and elucidate their intricate temporal dependence structure. 
Simulation studies are presented to evaluate the empirical performance of the SLS method. 
\end{abstract}


\end{frontmatter}
\section{Introduction}

Monitoring seismic data is crucial due to its significance in early warning systems, understanding fault systems, and facilitating post-earthquake analysis. The earthquake sequence on February 6, 2023 that occurred in southeast Turkey near the Syrian border, registering a magnitude of 7.8 and accompanied by numerous aftershocks in the thousands, underscores the significance of seismic data analysis~\citep{Turkey_Syria_earthquake}. By analyzing seismic data, scientists detected increased activity prior to the event, enabling timely warnings and evacuations. Monitoring also helps in comprehending fault behavior, improving forecasting models, and assessing damage after an earthquake.

Seismic data records earth motion over time, offering a temporal snapshot of subsurface structures. Since the early 20th century, seismic waves have been captured at high acquisition frequencies. In exploration geophysics, data is typically collected every 1 to 4 milliseconds (i.e., 1,000 to 250 Hz), resulting in massive data volumes \citep{yilmaz2001seismic}. 
This data influx offers valuable opportunities but also introduces significant computational challenges. The high data rate demands real-time processing, efficient storage, and fast retrieval systems to avoid bottlenecks. These requirements are especially critical for time-sensitive applications such as earthquake early warning systems, where delays in analysis can severely impact response effectiveness.
In practice, seismic data is treated as streaming time series, necessitating online algorithms for real-time collection and processing \citep{kropivnitskaya2018real}. Compared to simple random sampling, efficient data selection or sampling methods that automatically detect seismic events or ambient vibrations are preferred.

\begin{figure}[ht]
    \centering
    \includegraphics[width=1\linewidth]{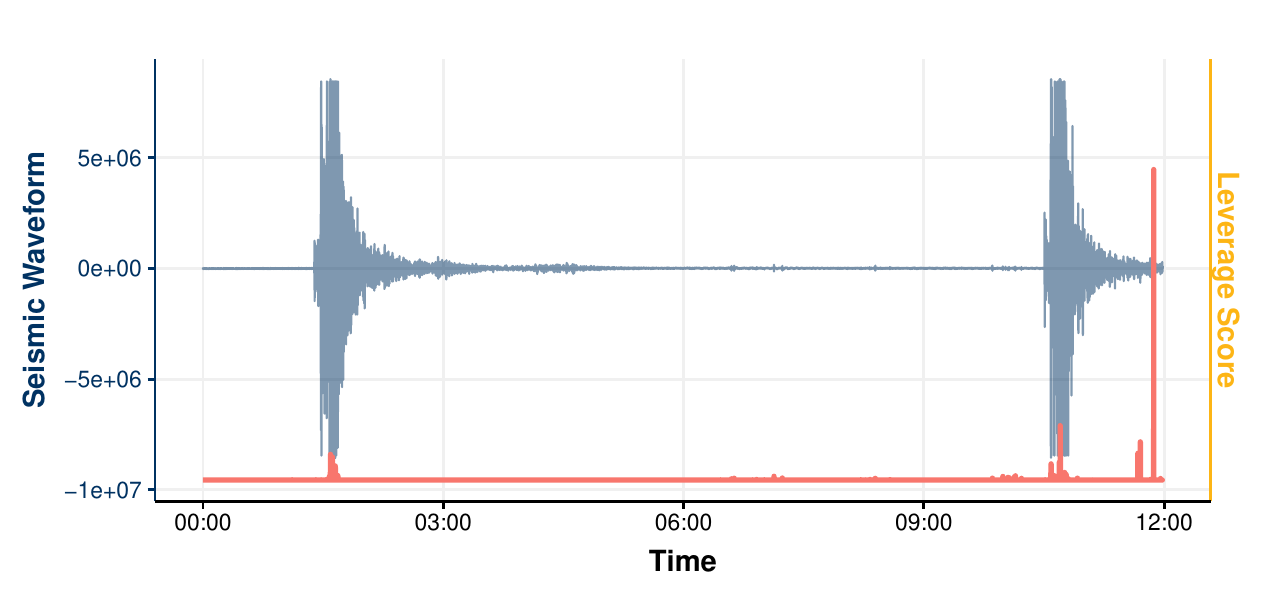}
    \caption{Seismogram of the 2023 Turkey–Syria earthquake doublet measure at the NL.TERZ
station (The Netherlands, Terziet) from 00:00 UTC to 12:00 UTC on February 6, 2023. The streaming leverage score $\tilde{h}_{ii}^{\text{AR}}$ of the seismic waves was displayed at the bottom of the figure.}\label{fig:earthquake}
\end{figure}

\begin{example}{\bf 2023 Turkey–Syria earthquake doublet.}\label{exmp1}  
On February 6, 2023, at 01:17 UTC, a moment magnitude (Mw) 7.8 earthquake struck southern and central Turkey, as well as northern and western Syria.
It was followed by another earthquake with a moment magnitude (Mw) of 7.7 at 10:24 UTC. This subsequent earthquake was centered 95 km (59 mi) north-northeast of the first one. The region experienced widespread damage, and the number of fatalities reached tens of thousands. This earthquake ranks as the fifth-deadliest of the 21st century. Figure~\ref{fig:earthquake} displays the seismogram of the earthquake doublet measure at the NL.TERZ station (The Netherlands, Terziet,~\cite{knmi:1993}). 
While developing a methodology for selecting or sampling streaming seismic signals and investigating its theoretical properties, we have a few tasks related to seismic data analysis in mind. Given our computational budget, can we observe changes in the dependency structure when seismic events occur? Can we utilize statistical inference tools to detect seismic events as soon as possible with a limited number of selected data points? Can we achieve better prediction accuracy when building a seismic model using a limited number of selected data points? We will provide answers by thoroughly applying our proposed method to seismic data analysis.
\end{example}
Seismic data can be effectively modeled as an autoregressive (AR) process, offering a tractable representation despite the simplifying assumption of i.i.d. errors in AR($p$) models. While seismic signals exhibit complex temporal dependencies that such models may fail to capture, linear models like AR($p$) remain valuable for several reasons.
First, they serve as first-order approximations that can offer interpretable insights into underlying physical mechanisms. In seismology, for example, linear models are closely tied to the analysis of focal mechanism~\citep{gephart1984improved}, such as fault orientation and slip, which motivates our use of these models. Second, although they may not model all nonlinear or long-range dependencies, the estimated coefficients from linear models can still reveal informative and interpretable structure that reflect key properties of the earthquake source and wave propagation \citep{takanami1991study,DARGAHINOUBARY1992381,kozin1988autoregressive}. 
Finally, given the constraints of real-time data processing, linear models offer computational efficiency and serve as a practical starting point for modeling streaming data. 
However, AR modeling is rarely applied in real-time seismic acquisition due to the high data volume, frequent sampling, and potential model dimensionality, which pose significant computational challenges, especially in resource-limited field settings. To enable timely and accurate parameter estimation, efficient data selection or sampling strategies are needed to reduce computational cost without sacrificing accuracy.

Streaming sensor data, particularly temporally correlated time series, presents major challenges for traditional estimation methods. As data volume grows rapidly, classical approaches like least squares (LS) and maximum likelihood estimation (MLE) become computationally infeasible, especially in real-time settings or on devices with limited memory that cannot store the entire dataset. In such contexts, data arrives continuously as a stream, not as a fixed dataset~\citep{babu2001continuous}, requiring scalable alternatives to conventional estimation techniques.

To address the computational challenges of real-time inference in streaming autoregressive time series, we propose a method called \textit{Sequential Leveraging Sampling} (SLS). SLS is an online data reduction technique designed to retain essential information from streaming data while reducing sample size. It features: (1) sequential block-wise sampling to account for temporal correlation, (2) scalability to high-frequency data, and (3) online parameter estimation. By selecting only the informative data points using a randomized leveraging strategy, SLS enables efficient, accurate inference under limited computing resources.
	

\textit{First}, SLS preserves temporal dependence in streaming data through sequential block sampling. By leveraging the autoregressive structure of the time series, SLS selects data blocks that retain the correlation among observations.

\textit{Second}, unlike recursive methods such as recursive least squares or the Kalman filter~\citep{ljung1983theory,harvey1990forecasting,guo1994further,young2012recursive}, which update estimates with every new observation, SLS is designed for high-frequency data streams. Recursive methods require frequent updates (e.g., $O(p^2)$ flops per iteration for recursive least squares~\citep{haykin1996adaptive}, where $p$ denotes the dimensionality of the parameter vector) and can be overwhelmed when data arrive faster than updates can be computed. For instance, even with efficient MATLAB implementations on a standard laptop, recursive updates become impractical when sample rates exceed 7 kHz. This is based on the recursive least squares estimation of AR($1$) time series with $512$ observations and $1000$ independent replicates.\footnote{This simulation study was conducted using MATLAB build-in function \textit{recursiveAR}~\citep{matlab} through a battery-powered laptop computer with 2.4 GHz Intel Core i5 CPU.} Real-world applications, such as medical imaging (15.7 kHz~\citep{yun2003high}), acoustic monitoring (200 kHz~\citep{wiggins2007high}), and seismic acquisition (1–1.5 MHz~\citep{hubbard2002mapping}) highlight the need for methods like SLS that select only a few representative blocks to reduce computational burden without sacrificing information.

\textit{Third}, SLS supports real-time analysis via an online, leverage-based data selection strategy that immediately evaluates and selects influential data points as they arrive. This selection does not require access to historical data, making SLS memory efficient and well-suited for streaming settings. Its sequential block sampling further enables scalable inference for high-volume streams, which is particularly novel and beneficial in seismic data analysis.

	
SLS employs a leverage-based, data-dependent sampling strategy to construct a compact ``sketch'' of the data stream, which serves as a surrogate for performing computation and statistical inference. This approach is detailed in Section~\ref{method} for linear autoregressive time series and extended to nonlinear settings in Section S.2 of~\cite{SupplementAOAS2094}.

\subsection*{Organization}
The remainder of the article is organized as follows. Section~\ref{relatedwork} reviews prior work on streaming data applications, recent advances in streaming analysis, and sampling methods for dependent data. Section~\ref{modl} introduces the autoregressive model, least squares estimation, and leveraging scores. Section~\ref{method} presents the proposed sequential leveraging sampling (SLS) method for streaming AR($p$) series. Section~\ref{Asymptotic} establishes uniform asymptotic normality and constructs fixed-width confidence intervals using the SLS block. 
Section~\ref{realdata} applies SLS to two seismic datasets. Section~\ref{s:Simulations} reports simulation results demonstrating the practical performance of SLS. Section~\ref{diss} concludes with a discussion of future directions. Technical details, additional simulations, and proofs are provided in the Supplementary Material~\cite{SupplementAOAS2094}. The Supplement~\cite{SupplementAOAS2094} also extends the SLS method to nonlinear autoregressive models with theoretical justification of its sample efficiency, as well as further extensions to nonlinear additive autoregressive models.
    
\section{Related work}\label{relatedwork}

The study of streaming data originally came from the field of computer science and engineering. 
 There is intensive research involving streaming data acquisition, storage, visualization, and information query, in communities of database, signal processing and pattern recognition of computer science and engineering ~\citep{fu2011review,papadimitriou2005streaming,woodruff2014data,garofalakis2016data}. 
 There is also a proliferation of literature on online analysis of streaming data, which usually requires real time analysis without the entire input data being available. Some representative examples are on methodology and software development~\citep{hoffman2010online,elhamifar2017subset}, and on online algorithms~\citep{keogh2001online,kossmann2002shooting}. 
The existing literature on statistical analysis of streaming data methods primarily revolves around the development and application of online iterative updating techniques for various models. Examples include \cite{luo2023real} for streaming clustered data, \cite{luo2023statistical} for streamed longitudinal data, \cite{luo2020renewable} for streaming data with generalized linear models, \cite{xue2020online} for streaming survival data, and \cite{fang2023online} for functional data online update, among others. 
To the best of our knowledge, the specific focus on sampling methods for streaming data is relatively lacking in the current body of literature.
		 \cite{xie:2019:online}  investigated online distributed estimation of a vector autoregression (VAR) model with the assistance of leverage score sampling.  \cite{eshragh2019lsar} applied leverage score sampling for the analysis of large-scale univariate time series. 
   Additionally, \cite{10.1214/23-AOAS1757} explored optimal sampling techniques for multivariate time series streams.

Recent randomized data reduction methods aim to address the computational challenges of analyzing massive datasets, including streaming data. Classical deterministic sampling methods, which require full data access and storage, are often impractical in streaming contexts due to high computational and memory costs.
As an alternative, randomized sketching via sampling has gained significant traction in large-scale data analysis~\citep{mahoney2011randomized}. This includes work on matrix approximation~\citep{drineas2006subspace,drineas2012fast,woodruff2014sketching}, least squares estimation~\citep{MaMY15,raskutti2016statistical}, compressed sensing~\citep{gilbert2007one}, and streaming applications such as anomaly detection~\citep{huang2015streaming} and network sampling~\citep{ahmed2014network,Gama2016RethinkingSA}.
In linear regression, \cite{drineas2006fast,drineas2012fast} introduced algorithmic leveraging, using empirical leverage scores for importance sampling and rescaling to reduce data size before computation. \cite{MaMY15} demonstrated its advantages over uniform sampling in big data settings, and \cite{raskutti2016statistical,ma2022asymptotic} further analyzed its statistical efficiency in generalized and sketched least squares problems.

Randomized sketching methods are generally designed for independent data and cannot be directly applied to streaming data, which are typically time-dependent. Non-uniform reduction techniques, such as leverage-based sampling, require full-sample calculations (e.g., leverage scores), which are infeasible in streaming settings due to the continuous nature of data. Moreover, statistical inference frameworks for streaming data and associated reduction algorithms remain underdeveloped.

		To determine the sample (block) size for data stream sampling, a sequential approach is preferred in the context of online analysis since the sample size should be data dependent rather than prespecified~\citep{lai1983fixed,barndorff1984effect}.
		 Our work adapts the idea of sequential expansion that keeps expanding the selected data block until the accumulated information reaches the pre-specified level.
		 

\section{Overview of the Problem and Preliminaries}\label{modl}
  We first consider a linear time series model for the streaming data $\{X_i\}_{i=-\infty}^{\infty}$, i.e. $p \ (\ge 1)$-th order autoregressive model (AR($p$)),
	\begin{equation}\label{eq:AR(p)}
	X_i=\beta_1 X_{i-1}+\beta_2 X_{i-2}+\ldots+\beta_p X_{i-p}+\varepsilon_i, 
	\end{equation}
	where $\boldsymbol {\beta} = (\beta_{1}, \ldots ,\beta_{p})^{T}$ is the unknown parameter vector, and the innovations $\{{\varepsilon_i}\}$ are assumed to be a sequence of independent and identically distributed (i.i.d.) random variables. 
	We assume that our observed data starts at $X_{1}, \ldots, X_{p}$, and the innovations $\{{\varepsilon_i}\}_{i=1}^{\infty}$ are independent of these starting values with $\E(\varepsilon_i)=0$ and $0<\V(\varepsilon_i) = \sigma^{2} <\infty$.
	For data observed up to time $n$, let ${\bf z}_{i}=[X_{i-1}, \ldots, X_{i-p}]^{T}$, and define the design matrix as
	\begin{equation}\label{designmatrix}
		\boldsymbol \Gamma_{n} =
		\left[ {\bf z}_{p+1}\ {\bf z}_{p+2} \ \cdots \ {\bf z}_{n} \right]^{T}.
	\end{equation}
	Using this notation, we can write the AR($p$) model observed up to time $n$ as
	${\bf x}_n=\boldsymbol \Gamma_{n} \boldsymbol \beta + \boldsymbol {\varepsilon}_n,$
	where ${\bf x}_n=[X_{p+1}, \ldots, X_{n}]^T$ and ${\boldsymbol \varepsilon}_n= [\varepsilon_{p+1}, \ldots, \varepsilon_{n}]^T$. 
	The least squares (LS) method fits the AR($p$) model by solving the optimization problem,
	 	\begin{equation}\label{eq:ls}
	   \min_{\boldsymbol \beta \in \mathbb R^p} ||{\bf x}_n - \boldsymbol \Gamma_n \boldsymbol \beta ||^2,
	 \end{equation}
 where $||\cdot||$ is the $\ell_2$ norm.
	For continuously observed streaming data, the ``sample size'' $n$ is infinite. Thus, the actual observed data can be arbitrarily large, making the exact LS solution $	\widehat{\boldsymbol \beta}_{n,LS}=   (\boldsymbol \Gamma_{n}^T \boldsymbol \Gamma_{n})^{\dagger}\boldsymbol \Gamma_{n}^T {\bf x}_n$ computationally challenging. Here $(\cdot)^{\dagger}$ is the Moore-Penrose inverse.

Sampling is a common strategy for reducing computational costs in large-scale problems. In SLS, we design a leverage-based sequential block sampling operator \( S \) to construct a sketched stream \( S{\bf x}_n \), enabling online least squares (LS) estimation on a smaller sub-problem. Instead of solving the full problem~\eqref{eq:ls} on the data stream \( ({\bf x}_n, \Gamma_n) \), which may be computationally infeasible, we perform LS estimation on the sketched pair \( (S{\bf x}_n, S\Gamma_n) \),

	\begin{equation}\label{eq:lss}
\widehat{\boldsymbol \beta}_{S}=  \arg \min_{\boldsymbol \beta \in \mathbb R^p} ||S{\bf x}_n -S \boldsymbol \Gamma_n \boldsymbol \beta ||^2,
\end{equation}
where $\widehat{\boldsymbol \beta}_{S}$ can be accurately estimated in a computationally efficient, memory-friendly, and fully online fashion, at the cost of losing other information due to discarding unused data.

\subsection{Statistical Leverage Scores for AR(p) Model}
For an AR($p$) model, the fitted values are expressed as $\widehat {\bf x}_n=\boldsymbol \Gamma_{n}\widehat{\boldsymbol \beta}_{n}={\bf H}_n{ \bf x}_n$, where ${\bf H}_n= \boldsymbol \Gamma_{n} (\boldsymbol \Gamma_{n}^T \boldsymbol \Gamma_{n})^{\dagger}\boldsymbol \Gamma_{n}^T$ is the so-called \emph{hat matrix}~\citep{HauTong1989}.
For ${\bf z}_i$ defined above, the $i^{th} $ diagonal element of ${\bf H}_n$,
\begin{equation}\label{leverage}
h_{ii}={\bf z}_i^T(\boldsymbol \Gamma_{n}^T \boldsymbol \Gamma_{n})^{\dagger}{\bf z}_i,
\end{equation}
is called the {\it statistical leverage} of the $i$th observation.  \cite{HauTong1989} showed that $h_{ii}$ may be interpreted as the amount of leverage or influence exerted on $\widehat X_{i}$ by $X_{i}$ and  $n h_{ii}$ is interpreted as the Mahalanobis distance between ${\bf z}_i$ and the zero mean vector. Furthermore, they established various properties of the hat matrix including that $0 \le h_{ii} \le 1$. Motivated by the work of \cite{drineas2006fast,drineas2012fast} and \cite{MaMY15} mentioned earlier, in Section~\ref{method} we will discuss how to use leverage scores to construct the ``sketch'' operator $S$ and then select the SLS block.

\section{Sequential Leveraging Sampling Method for Streaming AR($p$) Series}\label{method}

We propose to adaptively select only one block of consecutive data points, named {\it Sequential Leveraging Sampling} block, as a sketch of the streaming data. 
Under the streaming data setting, it is essential to design an efficient data selection method that can handle large volumes of streaming time series data and provide an accurate estimate of model parameters. 
 
 The SLS block consists of two key components: the starting point and sequential block size. First, we use a leverage-based randomized data selection method to obtain a starting point $X_{l}$ at time $l$ of the SLS block. Then, the starting point $X_{l}$ is expanded adaptively to form the SLS block ${\bf x_{\tau_c}}=[X_l,\ldots,X_{\tau_c}]^T$, where the stopping time $\tau_c$ is decided according to a sequential stopping rule.

When designing the SLS method, we account for key characteristics of streaming time series, namely, the data arrive continuously at discrete time points, are temporally correlated, and grow without bound. Accordingly,

\begin{itemize}
	\item Unlike in independent data settings where individual points are sampled, we sample blocks of consecutive observations to preserve temporal correlation.

	\item Blocks are selected to contain high-leverage points, as leverage-based sampling is generally more efficient than uniform sampling for both sample size and parameter estimation; see Section~S.2.3 of~\cite{SupplementAOAS2094} and~\cite{MaMY15}.

	\item A sequential stopping rule determines the block size, aiming for fixed-accuracy estimation. After selecting a start time, we expand the SLS block in real time, stopping once the accumulated information reaches a prespecified threshold.
\end{itemize}



	

Suppose we observe a streaming time series $\{X_{1}, X_2,\ldots\}$ following a streaming AR($p$) model in~(\ref{eq:AR(p)}). 
First, along with the data streaming through working memory, the starting point of the SLS block is selected according to the leverage score of the data point. In the online setting, we perform independent Bernoulli trials, where the success probability is proportional to the leverage score (capped at 1), to select the starting point of the SLS block. The idea of starting point selection can be traced back to~\cite{fan1962development}, where we adjust the calculation of the success probability to accommodate the streaming data selection. 	The rationale for matching the success probability with the leverage score is to exploit the fact that observations with a higher leverage score will have a higher probability of being selected as the starting point of the SLS block. We define the \textit{streaming leverage score} for an AR model as
\begin{equation} \label{streamlev}
    \tilde{h}_{ii}^{\text{AR}} := {\bf z}_i^{T}\hat{\Omega}_{n_0} {\bf z}_i\quad i\in\mathbb{Z},
\end{equation}
where we use $\hat{\Omega}_{n_0}$, an estimator of precision matrix $\Omega = (\E[\boldsymbol \Gamma_{n}^T \boldsymbol \Gamma_{n}])^{-1}$, to replace $(\boldsymbol \Gamma_{n}^T \boldsymbol \Gamma_{n})^{\dagger}$ in~\eqref{leverage} in online streaming calculation. The estimation $\hat{\Omega}_{n_0}$ is based on the pilot data of size $n_0$. This means that we can construct the leverage scores and the success probability even if the total number of data points is unknown. Moreover, the online calculation of the streaming leverage score is computationally efficient since we do not need to inverse the matrix for every incoming datum. The quality of online leverage score approximation and fast implementation can be found in ~\cite{mahoney2011randomized,drineas2012fast}.

\begin{algorithm}
	\caption{Sequential Leveraging Sampling Algorithm}
 \KwIn{Streaming time series $\{X_{1}, X_2,\ldots\}$}
 
 {\textbf{Initialization:} Order $p$ (selection based on BIC); 
 Precision matrix  $\hat{\Omega}_{n_0}$ (estimation based on pilot data $\{\mathbf{z}_1, \ldots, \mathbf{z}_{n_0}\}$); Information threshold $c>0$ (based on requirement on the width of confidence region).\\}
 Start the \textbf{online algorithm} for time $j \geq {n_0}+1$:
	\begin{algorithmic}[1]\label{Lev}
\STATE {\bf Starting value via independent Bernoulli trials}: For the subsequent data point $X_j$, draw an independent Bernoulli variable $B_j$ with success probability  $\min\{\tilde{h}_{jj},1\}$ ;
\\ \textbf{if} {\ \ \ $B_j=0$} \textbf{then} {$j\gets j+1$, go back to step 1;} \\
\textbf{else} {$B_j=1$ \Return starting time $l:=j$ and go to sequential expansion;}

\STATE {\bf Sequential expansion}: Expand the input stream, starting with $X_{l}$, to form a block of consecutive observations $\{X_{l},\ldots,X_{\tau_c}\}$, collected according to the sequential leveraging sampling rule $\tau_c = \inf\{ t \geq l:  \sum_{i=l}^{t} ||{\bf z}_i||^2 \geq c    \}$.
\STATE {\bf Least Squares Estimation on subproblem}: Calculate the LS estimator, $\widehat{\boldsymbol \beta}_{\tau_c} =  (\boldsymbol \Gamma_{\tau_c}^T \boldsymbol \Gamma_{\tau_c})^{\dagger}\boldsymbol \Gamma_{\tau_c}^T {\bf x_{\tau_c}}$, where $\boldsymbol \Gamma_{\tau_c}$ is defined in~(\ref{ssdesignmatrix}) and ${\bf x_{\tau_c}}=[X_l,\ldots,X_{\tau_c}]^T$.
\end{algorithmic}
\KwOut{SLS block ${\bf x_{\tau_c}}=[X_l,\ldots,X_{\tau_c}]^T$ and LS estimator $\widehat{\boldsymbol \beta}_{\tau_c}$.}
\end{algorithm}

Second, we suppose that time $l$ is the starting time determined by the independent Bernoulli trials and streaming leverage score $\tilde{h}_{ii}^{\text{AR}}$. 
Starting with $X_l$, we keep collecting consecutive data points to expand the SLS block until the sequential stopping rule is triggered at some time $\tau_c$: 
\begin{equation}\label{eq:stoppingtime}
	\tau_c = \inf\{ t \geq l:  \sum_{i=l}^{t} ||{\bf z}_i||^2 \geq c    \},
\end{equation}
where $c~(>0)$ is a prespecified constant called information threshold.
	Note that, if the ${\varepsilon}$'s are normally distributed,  $\sum_{i=l}^{t} ||{\bf z}_i||^2$ is the trace of the observed Fisher information matrix for the block $\{X_l,\ldots,X_{t}\}$. 
Accordingly, we define the design matrix of the SLS block as 
\begin{equation}\label{ssdesignmatrix}
	\boldsymbol \Gamma_{\tau_{c}}=
		\left[ {\bf z}_{l}\ \cdots \  {\bf z}_{s} \ \cdots \ {\bf z}_{\tau_{c}} \right]^{T},
\end{equation}
for stopping time $\tau_{c}$.

Finally, based on the SLS block ${\bf x_{\tau_c}}=[X_l,\ldots,X_{\tau_c}]^T$ and its design matrix $\boldsymbol \Gamma_{\tau_c}$, we get the least squares estimator of $\boldsymbol \beta$ on the subproblem, $\widehat{\boldsymbol \beta}_{\tau_c} =  (\boldsymbol \Gamma_{\tau_c}^T \boldsymbol \Gamma_{\tau_c})^{\dagger}\boldsymbol \Gamma_{\tau_c}^T {\bf x_{\tau_c}}$. 

To summarize, we propose Algorithm~\ref{Lev} 
to select the SLS block in an online fashion. In this algorithm,  there is no need to store previous data outside the SLS block or recompute any other quantities. The memory complexity of SLS is $O(p^2)$, which is independent of the sample size $n$.

\subsection{Hyperparameter Specification}\label{hyperparameter}
The remaining question is how to specify hyperparameters in SLS. We discuss several plausible specifications for the hyperparameters.

\smallskip
\noindent\textbf{Pilot Data.} The pilot data help determine several hyperparameters in the SLS. The size of the pilot data $n_0$, in the online setting, heavily depends on the availability of working memory. For the generic data selection scheme, the minimum size $n_0$ should guarantee the correct estimation of model order $p$ with high probability. We use the Bayesian information criterion (BIC;~\cite{schwarz1978estimating,hannan1979determination}) as a consistent estimation of model order $p$ for all compared data selection methods. Other popular model order selection techniques include the Akaike information criterion (AIC;~\cite{akaike1998information}), minimum description length (MDL;~\cite{rissanen1983universal}), or the autocorrelation and partial autocorrelation plots~\citep{box2011time}. For a detailed discussion on model selection and its uncertainty quantification, one can refer to the review~\cite{ding2018model} and references therein. 
Our simulation study (in Section \ref{pilot_sen}) indicates that the choice of pilot data size is quite insensitive with respect to the performance of the SLS method. 

\smallskip
\noindent\textbf{Streaming Leverage Score.}
	Data selection or sampling with probabilities proportional to leverage scores of the data matrix yields a high precision approximation to the original data matrix~\citep{drineas2008relative}. 
	However, leverage-based data selection is difficult to adapt naturally to data streams because leverage scores themselves are not easy to compute in a streaming setting.
	 The computation of exact leverage scores is not only expensive, but also impossible in the streaming setting because the leverage scores depend on all data points, including those that have not yet been observed in the data stream.

An important note is that we can approximate the leverage scores using eq.~\eqref{streamlev} to achieve a similar goal of data selection in a streaming setting. 
While selecting the model order $p$, we use pilot data to estimate the precision matrix $\Omega_{n_0}$ (computational complexity $O(n_0p^2)$) due to streaming setting and computational concerns. Using eq.~\eqref{streamlev},  the approximation of the leverage scores takes only $O(p^2)$ time, given the precision matrix $\hat\Omega_{n_0}$; whereas using eq.~\eqref{leverage}, the exact computation of $h_{ii}$, for $i \in [n]$,
requires $O(np^2)$ time.

\smallskip
\noindent\textbf{Information threshold $c$.}
The information threshold \( c \) represents the reciprocal of the SLS estimator’s accuracy, measured by the variance of its asymptotic distribution and prespecified by the user~\citep{lai1983fixed}. Remark~\ref{efficiency} relates \( c \) to the relative ``sample size''.

Let \(\mathcal{S}\subset \mathbb{Z}^+\) denote sampled indices, and \(\mathcal{S}_{SLS}\) the subset selected by SLS. Theorem 4.1 of~\cite{lai1983fixed} establishes that 
$\mathrm{P}_{\beta}\left\{\lim_{c \to \infty} |\mathcal{S}_{SLS}| c^{-1} = 1-\beta^2 \right\} = 1$ for each \(\beta \in (-1,1)\),
and for \(|\beta|=1\),
$
|\mathcal{S}_{SLS}| c^{-1/2} \xrightarrow{\mathcal{L}} \inf \left\{ t : \int_0^t W^2(s) ds = 1 \right\}$,
where \(\xrightarrow{\mathcal{L}}\) denotes convergence in law and \(W(t)\) is a standard Brownian motion. These asymptotics guide the choice of \( c \) or block size.


One practical guideline relates \( c \) to the confidence interval width for \(\boldsymbol{\beta}\). For example, in the AR(1) case, an approximate \((1-2\alpha)\)-level confidence interval is 
$\widehat{\beta}_{\tau_c} \pm c^{-1/2} \sigma \Phi^{-1}(1-\alpha)$,
where \(\Phi(\cdot)\) is the standard normal CDF. Proposition~S.4.4 of~\cite{SupplementAOAS2094} generalizes confidence region construction for AR(\(p\)) streams, with \( c \) incorporated via the stopping rule~\eqref{eq:stoppingtime}.

Alternatively, within memory limits, one can trial multiple \( c \) values and select the optimal based on criteria such as prediction error. However, this reduces computational efficiency. In practice, prior experience or pilot studies often provide reasonable choices for \( c \) (or stopping time).

\begin{figure}[tbh]
\centering   
\includegraphics[width=5in]{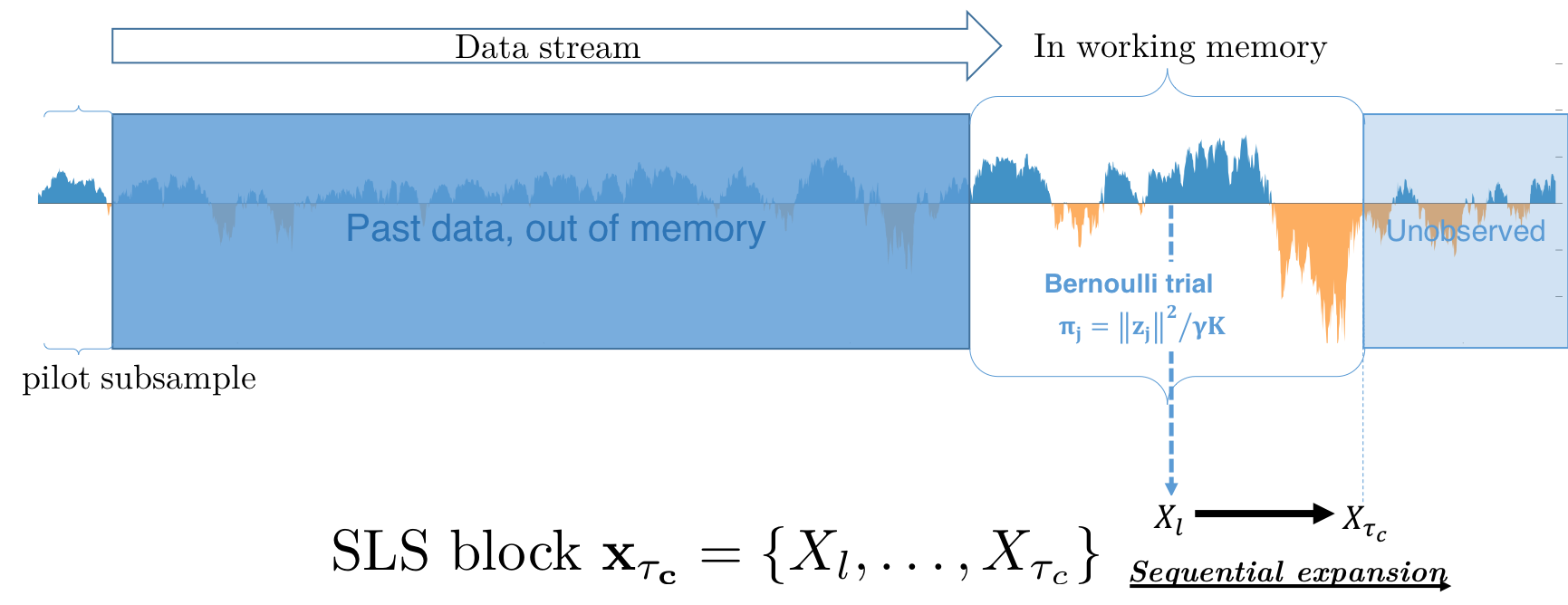}   \caption{   \textbf{An Illustration of SLS Algorithm~\ref{Lev}}. The sequence labeled with ``Data stream'' is the streaming time series we are observing. The SLS block, as a subset of the data points in the working memory,  with starting point $X_{l}$, selected according to leveraged-based independent Bernoulli trial (\ref{leverage}), and stopping point $X_{\tau_c}$ according to sequential stopping rule~(\ref{eq:stoppingtime}).  }\label{timeseriesfig:algo}
\end{figure}

\section{Theoretical Results for Linear AR Streams}\label{Asymptotic}
	 
   To provide a unified inference for streaming time series data coming from a $p$-th order autoregressive model (AR($p$)) with $p \ge 1$, we study the theoretical properties of our SLS method along the lines of~\cite{lai1983fixed},~\cite{galtchouk2011asymptotic} and related literature therebetween. 
 We establish a conditional uniform asymptotic normality result for the normalized least squares estimator based on the SLS method. We first present the main result for a streaming AR(1) series and then the results for the general AR($p$) model.

\subsection{Sequential Leveraging for AR(1) Streams} 

In the case of streaming first order autoregressive process AR($1$), we have $\mbf z_i =  X_{i-1}$. 
We follow the sequential leveraging algorithm in Section~\ref{method} to decide the starting time $l$  and the sequential stopping rule \eqref{eq:stoppingtime} of the SLS block.
Note that if $\varepsilon$'s are normally distributed, $\sum_{i=l}^t X_{i-1}^2$ is the observed Fisher information for the block $\{X_l,\ldots,X_{t}\}$. 
Recall that the estimator of $\beta$ based on the block $\{X_l, \cdots, X_{\tau_c}\}$ is given by 
$\widehat{\beta}_{\tau_c}= \sum_{j=l}^{\tau_c} X_{j-1}X_j / \sum_{j=l}^{\tau_c} X_{j-1}^2$. We now state the main theorem.


\begin{theorem}\label{normalityar1}
If $\varepsilon_{l},\varepsilon_{l+1}, \ldots, $ are i.i.d. with mean $0$ and variance $\sigma^2$, and the sequence $\{\varepsilon_{i}: i \ge l\}$ is independent of $X_{l-1}$ (defined in Algorithm~\ref{Lev}), then
\begin{equation}\label{ar1clt}
\lim_{c\to \infty}  \sup_{|\beta|\le 1} \sup_{x \in \mathbb{R}} \left|\mathrm{P}_{\beta} \left[ \sqrt{\sum_{i=l}^{\tau_c} X_{i-1}^2}  \left( \widehat{\beta}_{\tau_c} -   \beta \right) \le x \right]-\Phi(x/\sigma) \right|=0,
\end{equation}
where $\mathrm{P}_{\beta}$ is the conditional probability measure defined in Lemma~S.4.2 of~\cite{SupplementAOAS2094}.
\end{theorem}

\bigskip
The proof of Theorem~\ref{normalityar1} has been relegated to the Appendix. The proof of Theorem~\ref{normalityar1} is along the line of Theorem 2.1 of~\cite{lai1983fixed} with appropriate conditional probability measures.  Theorem~\ref{normalityar1} provides a uniform asymptotic normality
result given the starting time $l$. That is, based on the uniformity result of~(\ref{ar1clt}), the sequential leveraging sampling establishes a unified approach for the estimation of $\beta$ regardless of whether $|\beta| < 1$ or  $|\beta|= 1$.


\begin{remark}\label{efficiency}
 To quantify the loss in inferential efficiency due to sampling, we consider an asymptotic analysis comparing SLS with ``full sample'' time series of size $n$ in terms of their relative efficiency. The asymptotic relative efficiency of SLS estimator,  $\widehat{\beta}_{\tau_c}$, and ``full sample'' LS estimator, $\widehat{\beta}_{n}$, is the ratio of their asymptotic variances, 
\[
e(\widehat{\beta}_{\tau_c}, \widehat{\beta}_{n}) = \frac{\V{(\widehat{\beta}_{n})}} {\V{(\widehat{\beta}_{\tau_c})}}= \frac{\sigma^2 /n}{\sigma^2 {(\sum_{i=l}^{\tau_c} X_{i-1}^2)}^{-1} } = \frac{\sum_{i=l}^{\tau_c} X_{i-1}^2}{n} \approx \frac{c}{n},
\]
for fixed $\beta\in(-1,1)$. 

We interpret the relative efficiency as the effective ``sample size'' of SLS required to achieve the certainty of the full sample. However, in streaming settings with infinite data, one may instead select the information criterion \( c \) to target a desired accuracy. The estimation accuracy of the SLS estimator for \( \beta \), measured by the variance of its asymptotic distribution, is approximately a small constant \( c^{-1} \)~\citep{lai1983fixed}.

\end{remark}


\subsection{Sequential Leveraging for AR(p) Streams}\label{sec:arp}

We next consider the uniform asymptotic normality properties of SLS least square estimator $\widehat{\boldsymbol \beta}_{\tau_{c}}$ that is eligible for both the stable and the unstable AR($p$) streams. 

\begin{theorem}\label{thm:arpclt}
	Let $\hat{\boldsymbol \beta}_{\tau_c}$ be the least squares estimate of $\boldsymbol \beta$ based on design matrix~(\ref{ssdesignmatrix}), with $\tau_c$ defined in~(\ref{eq:stoppingtime}) and ${\bf V}_{\tau_{c},\,\boldsymbol \beta} = (\boldsymbol \Gamma_{\tau_{c}}^T \boldsymbol \Gamma_{\tau_{c}})^{1/2}(\widehat{\boldsymbol \beta}_{\tau_{c}} -  \boldsymbol \beta)$. Assume that $\boldsymbol \beta$ satisfies Conditions $1$, $2$ and $3$ in S.1.4 of~\cite{SupplementAOAS2094}. If $\varepsilon_{l},\varepsilon_{l+1}, \ldots, $ are i.i.d. with mean $0$ and variance $\sigma^2$, and the sequence $\{\varepsilon_{i}: i \ge l\}$ is independent of the starting state ${\bf z}_{l}=[X_{l-1}, \ldots, X_{l-p}]^{T}$ defined in Algorithm~\ref{Lev}, then 
	\begin{equation}\label{arpclt1}
\lim_{c \rightarrow \infty}  \sup_{\boldsymbol \beta \in K}  \sup_{\boldsymbol{x}  \in \mathbb{R}^p} \left|  \mbf{F}_{{\bf V}_{\tau_{c},\,\boldsymbol \beta}}(\boldsymbol{x})   -   \mbf{\Phi}(\boldsymbol{x}/\sigma)\right|=0,
	\end{equation}
	for any compact set $K \subset \tilde{\Lambda}_p$.
\end{theorem}


The proof of Theorem~\ref{thm:arpclt} has been relegated to Supplementary Material~\cite{SupplementAOAS2094}. With the result of Theorem~\ref{thm:arpclt}, the following theorem can be proved easily and is thus omitted.
\begin{theorem}\label{thm:arpchi}
	Let $\hat{\boldsymbol \beta}_{\tau_c}$ be the least squares estimate of $\boldsymbol \beta$ based on design matrix~(\ref{ssdesignmatrix}) and $\tau_c$ be defined in~(\ref{eq:stoppingtime}).  Assume that $\boldsymbol \beta$ satisfies Conditions $1$, $2$ and $3$ in S.1.4 of~\cite{SupplementAOAS2094}. If $\varepsilon_{l},\varepsilon_{l+1}, \ldots, $ are i.i.d. with mean $0$ and variance $\sigma^2$, and the sequence $\{\varepsilon_{i}: i \ge l\}$ is independent of the starting state ${\bf z}_{l}=[X_{l-1}, \ldots, X_{l-p}]^{T}$ defined in Algorithm~\ref{Lev}, then 
	\begin{equation}\label{arpclt2}
		\frac{1}{\sigma^2} (\widehat{\boldsymbol \beta}_{\tau_c} -  \boldsymbol \beta)^T(\boldsymbol \Gamma_{\tau_c}^{T} \boldsymbol \Gamma_{\tau_c})(\widehat{\boldsymbol \beta}_{\tau_c} -  \boldsymbol \beta) \xrightarrow{}  \chi^2_p ,\mbox{    }\text{as } c \to \infty,
	\end{equation}
	uniformly in $\boldsymbol \beta \in K$ for any compact set $K \subset \tilde{\Lambda}_p$.
\end{theorem}

We are now able to connect Theorem~\ref{thm:arpchi} to a fixed width confidence region based on $\hat{\boldsymbol \beta}_{\tau_c}$. For a data block $\boldsymbol\Gamma_{\nu}=\left[ {\bf z}_{l}\ \cdots  \ {\bf z}_{\nu} \right]^{T}$ of size $\nu-l+1$ and a constant $d\in \mathbb{R}^{+}$, we define the ellipsoid that has a fixed length of the major axis of $2d$
\begin{equation}
	\boldsymbol R_{\nu}(d) := \{ \boldsymbol \beta: (\boldsymbol \beta - \widehat{\boldsymbol \beta}_{\nu} )^T(\boldsymbol \Gamma_{\nu}^{T} \boldsymbol \Gamma_{\nu})(\boldsymbol \beta - \widehat{\boldsymbol \beta}_{\nu}) \leq d^2 \tr(\boldsymbol \Gamma_{\nu}^{T} \boldsymbol \Gamma_{\nu}) \},
	\end{equation}
where $\tr(\boldsymbol \Gamma_{\nu}^{T} \boldsymbol \Gamma_{\nu})$ is the trace of observed Fisher information matrix $ \boldsymbol \Gamma_{\nu}^T \boldsymbol \Gamma_{\nu}$. 
Given any significance level $\alpha \in (0,1)$, we let $\nu_0$ be determined by
\begin{equation}\label{n0}
	{\nu}_0(d) := \lceil{\sigma^2 a^2/[d^2 \tr(\E_{\boldsymbol{\beta} }(\boldsymbol \Gamma_{\nu}^T \boldsymbol \Gamma_{\nu} ))]}\rceil
	\end{equation}
where $\lceil{\cdot}\rceil$ is the ceiling function, which maps a real number to the least succeeding integer, and $a^2$ satisfies $\mathrm{P} [\chi^2_p \leq a^2] = 1- \alpha$. From \eqref{arpclt2}, for a stable series, i.e. $\boldsymbol \beta \in \Lambda_p$, we have
\begin{equation}\label{tau0}
	\lim_{d\to0} \mathrm{P} \left( \boldsymbol \beta \in \boldsymbol R_{{\nu}_0(d)}\right) = 1- \alpha.
\end{equation}
The result in~\eqref{tau0} suggests that the block size $\nu_0-l+1$  yields an ellipsoidal confidence region of fixed size for small width values of $d$ at specified coverage probability~\citep{sriram2001fixed}.
	
   Thus, to implement the sequential stopping rule in~(\ref{eq:stoppingtime}) and thereby determine the practical stopping time $\tau_c(d)$, we replace $\nu_0(d)$ in~(\ref{n0}) with $c = \hat{\sigma}^2 a^2 / d^2$, so that the resulting ellipsoidal confidence region has the desired fixed size and coverage probability.
    An estimator of $\sigma^2$ is $\hat{\sigma}^2_{n} = \frac{1}{n}||{\bf x}_{n} - \boldsymbol \Gamma_{n} \boldsymbol \beta ||^2$ with a block of size $n$, and $a^2$ satisfies $P[\chi_p^2 \leq a^2] = 1 - \alpha$.
    It leads to 
 \begin{equation}\label{arpclt}
		\lim_{d\to 0} 
		 \mathrm{P} \left({\boldsymbol \beta} \in \boldsymbol R_{\tau_c(d)}\right) = 1- \alpha,
	\end{equation}
 where the details are discussed via Proposition~S.4.4 of~\cite{SupplementAOAS2094}.


\section{Seismic Data Analysis}\label{realdata}
In this section, we analyze  the performance of the SLS method on two seismic data streams of different scales. The 2023 Turkey-Syria earthquakes (\ref{Turkey}) represent a large-scale, major earthquake that lasted for 12 hours and generated strong seismic signals. On the other hand, the Oklahoma Seismic Data (\ref{Oklahoma}) represents a micro-scale seismic event that occurred within seconds. 
In both cases, we compared the performance of the proposed SLS method to ``Uniform Sequential Sampling'' ({\qcr Uniform}), which refers to sequential
sampling with starting points chosen randomly with equal probability (i.e., Bernoulli trials
with equal probability), in terms of AR parameter estimation and prediction accuracy. The results showed that the SLS method outperformed naive sampling in handling streaming seismic signals and providing more reliable results for online seismic data analysis.

\subsection{2023 Turkey–Syria earthquakes}\label{Turkey}
We applied the SLS method to the 12-hour long (from 23:59:47 GMT on February 05, 2023, to 11:59:01 GMT on February 06, 2023) seismic stream from the 2023 Turkey-Syria earthquakes~\citep{knmi:1993}, where the sampling frequency was 40 Hz (40 records per second). The sampling and monitoring process was conducted in an online fashion using a total of 1.726 million seismic records, as depicted in Figure~\ref{fig:earthquake}.
Our primary aim is to evaluate the ability of the selected SLS blocks to capture changes in dependency structure during the monitoring process. This will provide us with a quick understanding of the ground motion generation mechanism~\citep{chang1982arma,DARGAHINOUBARY1992381}. Additionally, our secondary goal is to detect seismic events as quickly as possible using the statistical inference tool developed for SLS when monitoring the seismic stream.

During the SLS initialization, we utilized the first 10 minutes of seismic data, consisting of 24,000 seismic records, to estimate the order of the AR model $p$, using the Bayesian Information Criterion (BIC). We also calculated the precision matrix $\hat{\Omega}_{n_0}$. The threshold for information, denoted as $c$, was selected to be approximately equivalent to the information contained in a 1-minute duration of a stationary seismic stream without any earthquake events.

\begin{figure}
    \centering
    \begin{minipage}[t]{0.5\textwidth} 
    \centering
    \label{tab:ar_coefficients} 
    \begin{tabular}{lcccccc}
        \toprule
        & Pilot Data & Before Earthquake & Mw 7.8 Earthquake & Mw 7.7 Earthquake & Aftershock & Full Data \\
        \midrule
        lag1 & 1.1376 & 1.0829 & 1.9861 & 1.3207 & 0.9969&0.8228 \\
        lag2 & -1.3608 & -1.3712 & -0.9861 & -0.3208 & 0.9969&-0.5195 \\
        \bottomrule
    \end{tabular}
    \end{minipage}\hfill 
    \begin{minipage}[t]{0.5\textwidth} 
    \centering
\includegraphics[width=0.95\linewidth]{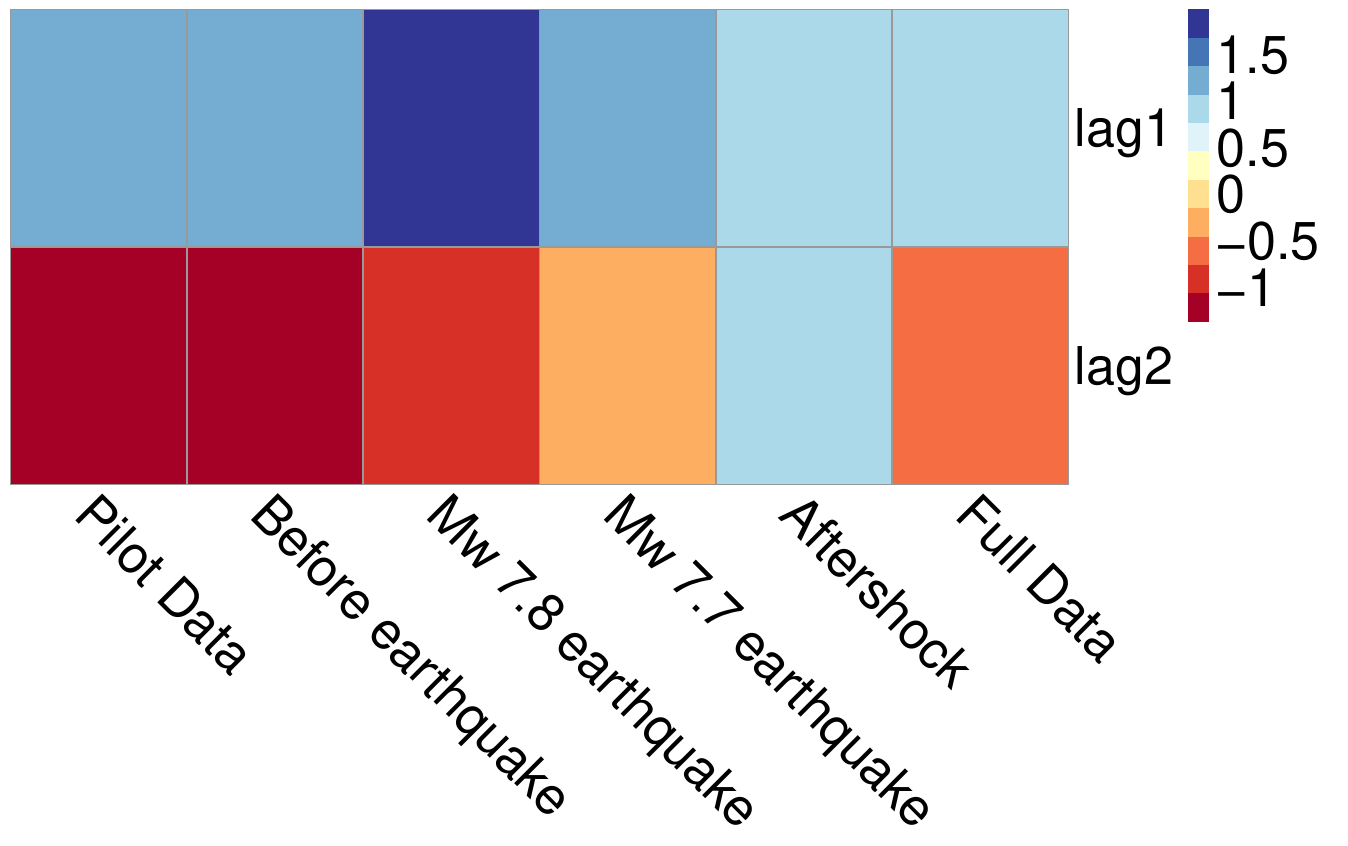} 
    \end{minipage}
    \caption{The first two lags of the estimated AR model coefficient $\widehat{\boldsymbol \beta}_{\tau_c}$ and heatmap visualization at different time points corresponding to different SLS blocks.}
    \label{fig:turkey_beta}
\end{figure}

In total, we selected 3,159 SLS blocks at various time points and with varying block lengths, providing a sketch of the underlying 24-hour seismic stream.
Each of these 3,159 SLS blocks corresponds to a sampled time series segment. For each block, we apply Ordinary Least Squares (OLS) based on the sampled segment (as described in Step 3 of Algorithm 1) to obtain an individual estimate.
In Figure~\ref{fig:turkey_beta}, we report and visualize the first two lags of the estimated AR model parameters, $\widehat{\boldsymbol \beta}_{\tau_c}$, at different time points corresponding to selected SLS blocks (5 out of the 3,159), along with the estimate based on the full sample.
Although the pilot estimation suggested an AR order of 14, some SLS blocks yield lower-dimensional LS estimates due to shifts in temporal dependency, often signaling earthquake events. To ensure consistency, we visualize only the first two lag coefficients in Figure~\ref{fig:turkey_beta}.
We observed that the coefficients from the pilot estimation and before the earthquake (first two columns) are similar.
However, in the third and fourth columns, we noticed an increase in the magnitude of the first lag coefficient and a decrease in magnitude for the two major earthquakes. Following the occurrence of the major earthquakes, we observed a change in the coefficients for both lags, with both becoming positive (the last column). This differs from the coefficients observed before the earthquakes.

\begin{figure}[h]
    \centering
    \includegraphics[width=1\linewidth]{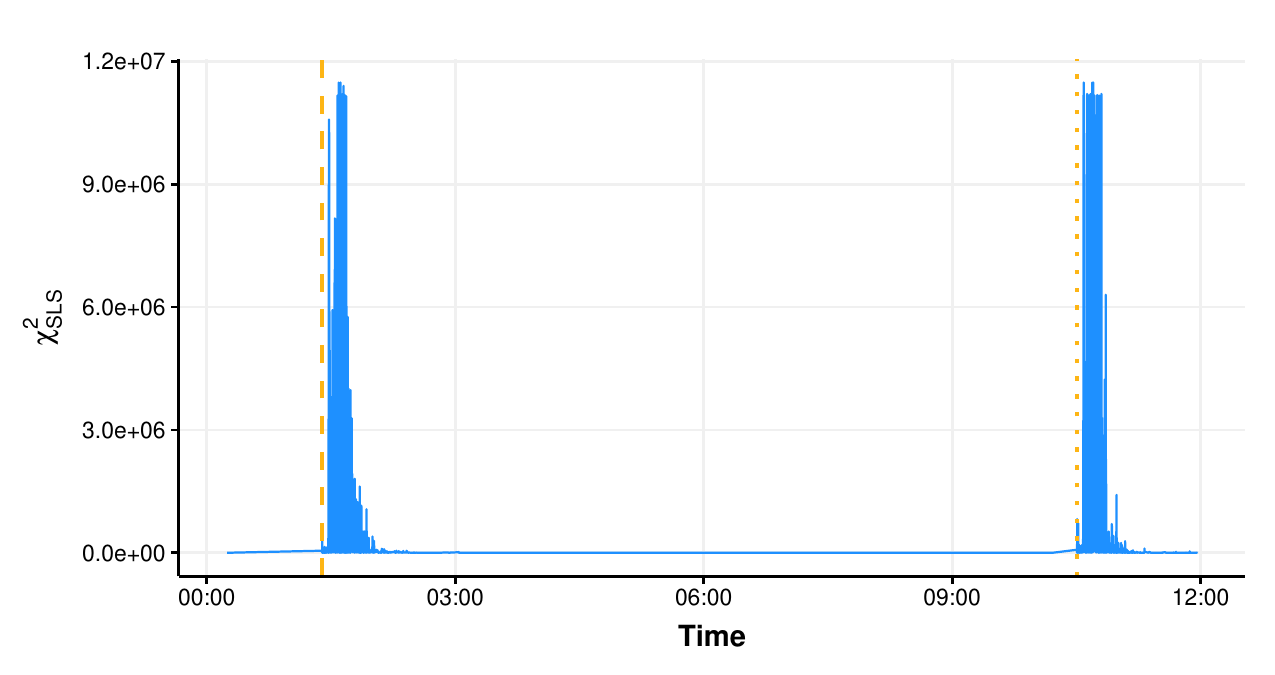}
    \caption{The $\hat\chi^2_{\text{SLS}}$ statistics based on the SLS blocks utilized for monitoring the deviation from the stationary seismic state. The dashed vertical lines serve as visual markers, highlighting the specific points in time when these significant seismic events occurred. }
    \label{fig:turkey_chisq}
\end{figure}

We now examine the ability of the proposed SLS method to detect earthquake events even under limited access to seismic records.
While the assumption of stationarity is made in theoretical analysis, it is important to note that seismic events can cause changes in the intrinsic time series model. However, we can still apply our SLS method in practice to detect and understand seismic events. This allows us to provide valuable insights and a better understanding of these events.
In Theorem~\ref{thm:arpchi}, we derived the asymptotic distribution of the pivot, which follows a $\chi_p^2$ distribution. It serves as a guide for detecting possible deviations from the stationary seismic state where no earthquake occurs.
 We calculate the $\hat\chi^2_{\text{SLS}}$ statistic from the SLS block,
\[\hat{\chi}^2_{\text{SLS}}=\frac{1}{\hat{\sigma}_0^2} (\widehat{\boldsymbol \beta}_{\tau_c} -  \widehat{\boldsymbol \beta}_{0})^T(\boldsymbol \Gamma_{\tau_c}^{T} \boldsymbol \Gamma_{\tau_c})(\widehat{\boldsymbol \beta}_{\tau_c} -   \widehat{\boldsymbol \beta}_{0}), \]
where $\hat{\sigma}_0$ and $\widehat{\boldsymbol \beta}_{0}$ are the pilot estimators of innovation variance and AR model coefficient based on the pilot data. In Figure~\ref{fig:turkey_chisq}, we observed that only based on sampled SLS blocks (3159 SLS blocks out of 1.726 million seismic records), we are able to detect the onset of the two major earthquakes. 
The $\hat\chi^2_{\text{SLS}}$ offers a real-time normalized measure that provides a quantifiable metric for theoretical reliability in monitoring and detecting any possible deviations from the expected seismic behavior.

The SLS method also serves as a high-frequency monitoring and hazard alert tool. By calculating the streaming leverage score for each data point, it may detect earthquake events and provide timely warnings. This feature is an advantageous by-product of the SLS method, ensuring efficient monitoring and alert capabilities.
The bottom part of Figure~\ref{fig:earthquake} displays the streaming leverage score $\tilde{h}_{ii}^{\text{AR}}$ of each seismic record. 
We observed that the spikes in the streaming leverage score coincide with the major seismic events.
However, the theoretical behavior of the streaming leverage score and its utilization as a tool for anomaly detection are beyond the scope of our article.

\begin{figure}[htb]
	\centering    \includegraphics[width=5.5in]{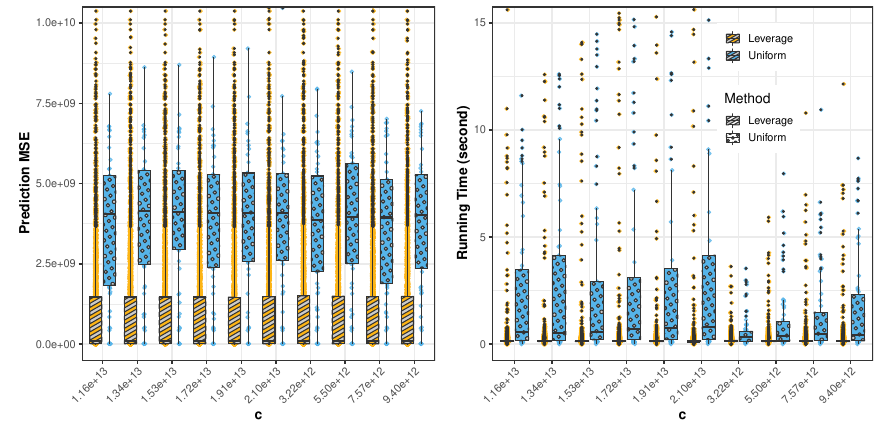}   \caption{Turkey Seismic Data. Left: Boxplots of \textbf{prediction MSE} of AR($14$) model on the test data for SLS ({\qcr Leverage}) and {\qcr Uniform}  at different information threshold $c$ levels. Right: Boxplots of \textbf{running time }of the SLS ({\qcr Leverage}) and {\qcr Uniform} at different information threshold $c$ levels.
	 }\label{turkeyplot}
\end{figure}
The comparative performance of the proposed SLS method and the naive  {\qcr Uniform} sampling method is illustrated in Figure~\ref{turkeyplot}, where the boxplots of prediction MSE (left) and running time (right) are displayed across varying information thresholds $c$. For this evaluation, the test data were selected from a seismic event corresponding to the second major shockwave of the 2023 Turkey–Syria earthquakes, occurring at 2023-02-06 10:38:01 GMT. As shown, the SLS method consistently achieves substantially lower prediction MSE than the Uniform method at all threshold levels, indicating superior predictive accuracy. In addition, the SLS method demonstrates significantly reduced computational cost, as evidenced by its consistently shorter running time compared to the Uniform method. These results highlight the efficiency and effectiveness of SLS for large-scale seismic data analysis.

 \subsection{Oklahoma Seismic Data} \label{Oklahoma}
\begin{figure}[hb]
	\centering    \includegraphics[width=5in]{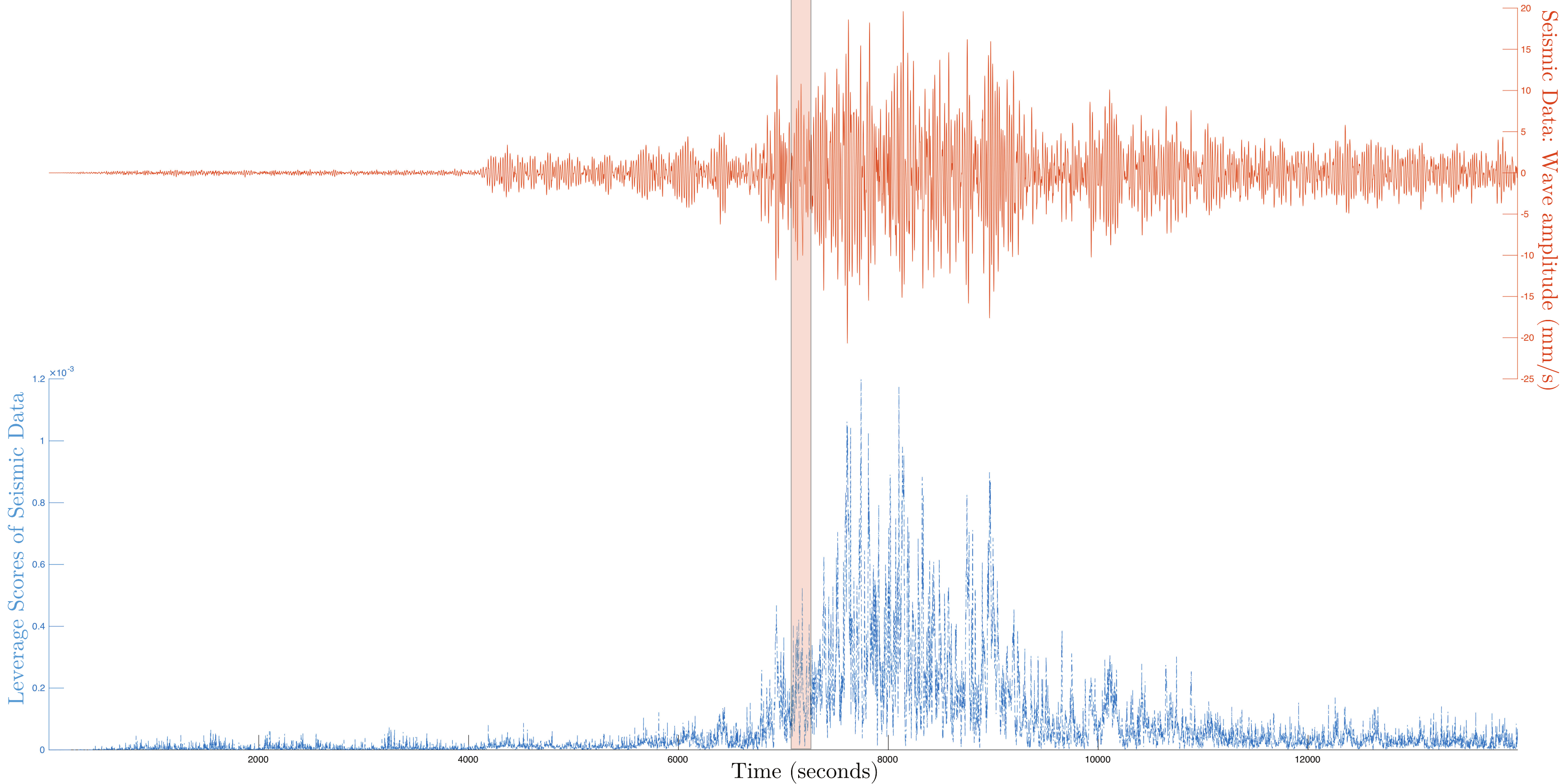}   \caption{   \textbf{Oklahoma seismic stream that has one seismic event} Top: Scatter plot of the seismic data (solid line). Bottom: Corresponding leverage scores for the AR($4$) model (dash-dot line). As an example, the highlighted area indicates one of the $100$ SLS blocks. }\label{seis}
\end{figure}

The seismic data analyzed in this section were well-recorded earthquake sequences (wave amplitude, $mm/s$) in Oklahoma that were collected on October 26, 2014. 
This dataset represents a small-scale earthquake with a short duration (in seconds).
We refer to~\cite{seisdata} for details of the seismic data. The total sample size for the earthquake sequence is $16,000$. The seismic data is modeled as an AR($p$) process with $p$ chosen to be $4$ ($p=4$) based on the pilot analysis of size $200$. From Figure~\ref{seis} , the plot of leverage scores of the seismic data (lower part, dash-dot line), we observe that the one seismic event is clearly illustrated by the picks. The starting point sampling probability constructed based on leverage scores will boost the sequential leveraging method to capture the seismic events.

In our analysis, we treated the seismic data as a streaming time series and divided the data into a training set and a test set. The training set, which contains the first $14,000$ time points, is used to estimate the model parameters using the SLS method and {Uniform} method. 
The test set, which contains the last $2000$ time points, is used to evaluate the estimation accuracy via the prediction MSE.

\begin{figure}[htb]
	\centering    \includegraphics[width=5in]{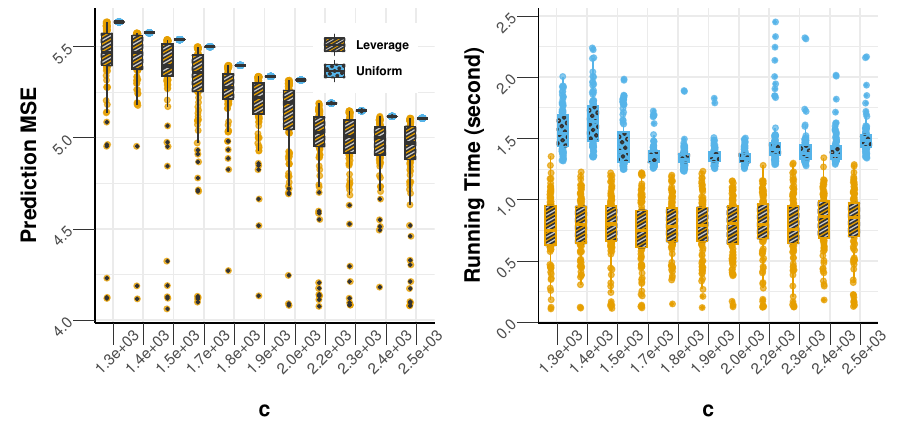}   \caption{Oklahoma Seismic Data. Left: Boxplots of \textbf{prediction MSE} of AR($4$) model on the test data for SLS ({\qcr Leverage}) and {\qcr Uniform}  at different information threshold $c$ levels. Right: Boxplots of \textbf{running time }of the SLS ({\qcr Leverage}) and {\qcr Uniform} at different information threshold $c$ levels.
	 }\label{seisplot}

\end{figure}

We perform the SLS method for the AR($4$) model to get sketches of the seismic stream with pilot data of size $n_0=200$. There are $11$ different values, ranging from $1.2\times 10^3$ to $2.5\times 10^{3}$, of information threshold $c$ evaluated, and each setting of $c$ is calculated with $n_{rep}=100$ replicates.
The SLS results are demonstrated in Figure~\ref{seisplot} comparing with the {Uniform} method, where the boxplots of prediction MSE (left) and running time (right) are plotted at different information threshold $c$.  We see that our SLS method outperforms the {Uniform} method with consistently lower prediction MSE than those of the {Uniform} method at all information threshold levels.

Figures~\ref{turkeyplot} and~\ref{seisplot} demonstrate that the information threshold parameter $c$ has minimal effect on the computational time of the SLS method. This phenomenon can be attributed to two primary factors. First, for a fixed $c$ value, the SLS method tends to select shorter blocks than the Uniform method while achieving the same information threshold. As a result, SLS requires fewer computations, leading to reduced overall runtime. Second, although the block size varies with different $c$ values for both methods, the magnitude of this variation is relatively small. The resulting changes in block size are not sufficient to substantially alter the computational bottleneck, which is predominantly influenced by block size. Therefore, computation time remains relatively stable across different $c$ values for both methods, with SLS consistently exhibiting superior efficiency compared to the Uniform method.



\section{Simulation Studies}\label{s:Simulations}
In this section, we demonstrate the empirical performance of the SLS method for linear AR($1$) streaming data and nonlinear autoregressive (NLAR) streams with known function forms.
We have implemented the SLS method presented in Section~\ref{method} using R Version 3.6.2.
Comprehensive simulation studies are presented for parameter estimation, running time, block size, and sensitivity analysis using synthetic streaming AR(1)  time series with various parameter settings. Additional simulation studies using synthetic streaming AR(1), AR(2), Nonlinear AR and Nonlinear Additive AR time series are presented in the Supplementary Material~\cite{SupplementAOAS2094}.

%
\subsection{AR(1) Series} 
\begin{figure}
       \caption{ \label{tbl:AR1_t4}Boxplot comparing MSE of parameter estimation, running time and block size for AR($1$) streams with varying information threshold $c$, observed with Sequential Leveraging Sampling and other two benchmark methods, on $100$ independent replications.}

         \centering
\begin{tabular}{
  >{\centering\arraybackslash}m{0.5cm}@{\hspace{2mm}}
  >{\centering\arraybackslash}m{0.8cm}@{\hspace{0.1mm}}
  >{\centering\arraybackslash}m{40mm}@{\hspace{0.1mm}}
  >{\centering\arraybackslash}m{40mm}@{\hspace{0.1mm}}
  >{\centering\arraybackslash}m{40mm}
}
           \toprule
            $\beta$ & $c$ & MSE & Running Time & Block Size \\
            \midrule
           \scriptsize $-0.3\ $&\scriptsize$500$ &\includegraphics[width=3.2cm]{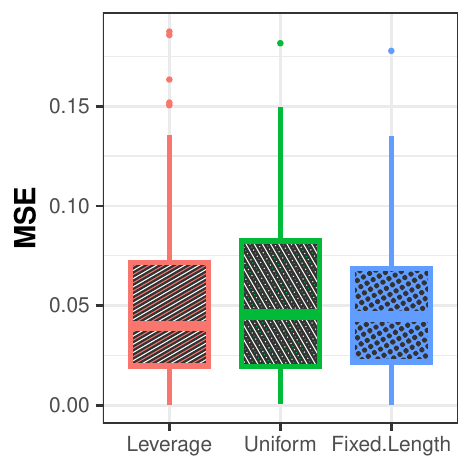}  & \includegraphics[width=3.2cm]{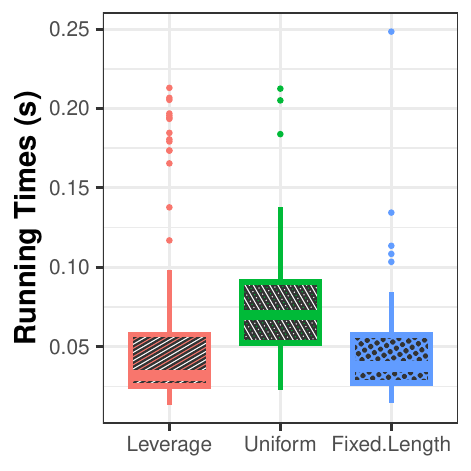} & \includegraphics[width=4.6cm]{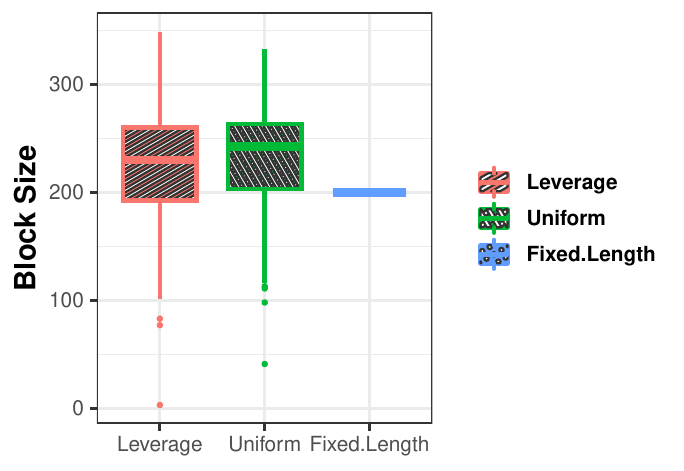}  \\
          \scriptsize $-0.5\ $&\scriptsize$600$ &\includegraphics[width=3.2cm]{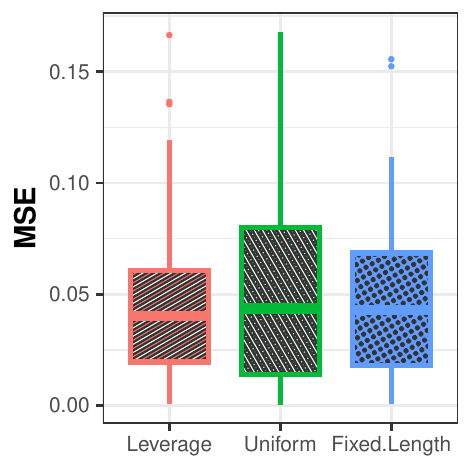}  & \includegraphics[width=3.2cm]{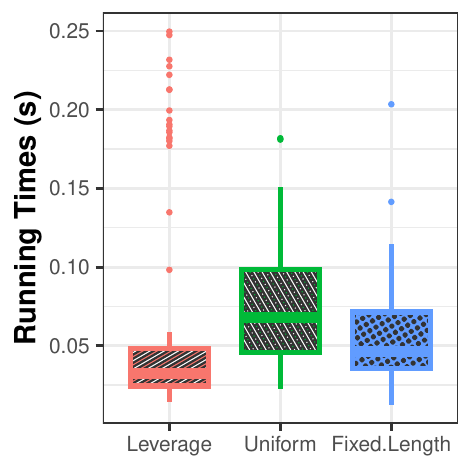}  &  \includegraphics[width=4.6cm]{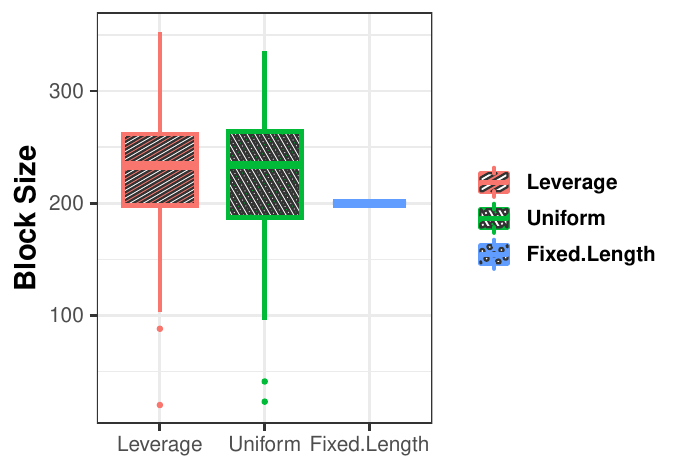} \\
          \scriptsize $-0.9\ $&\scriptsize$3000$ &\includegraphics[width=3.2cm]{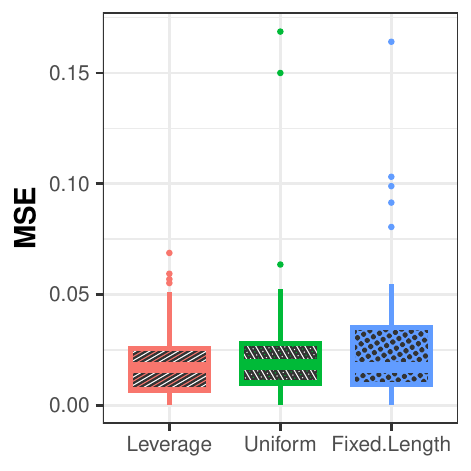} & \includegraphics[width=3.2cm]{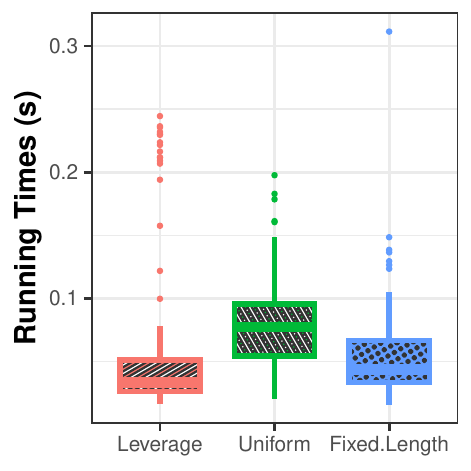}  & \includegraphics[width=4.6cm]{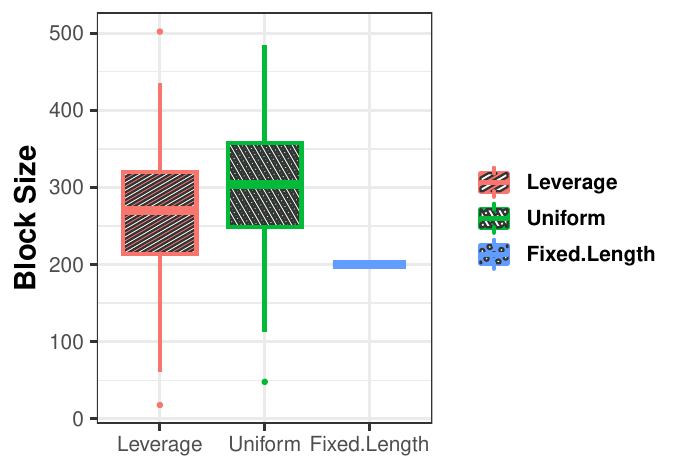} \\
             \scriptsize  0.99&\scriptsize$20000$ &\includegraphics[width=3.2cm]{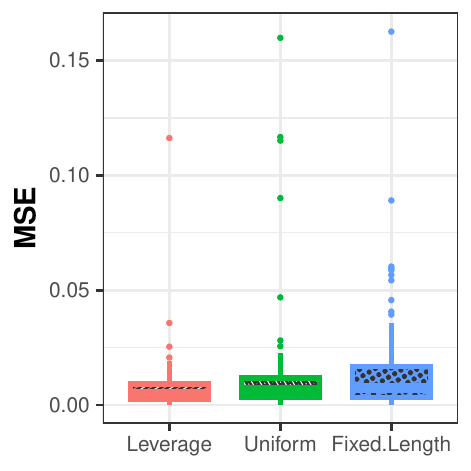}  & \includegraphics[width=3.2cm]{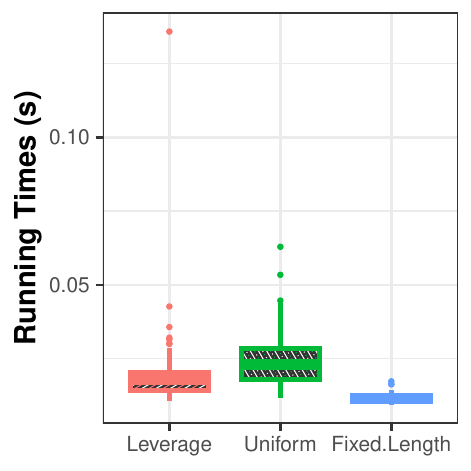}  &  \includegraphics[width=4.6cm]{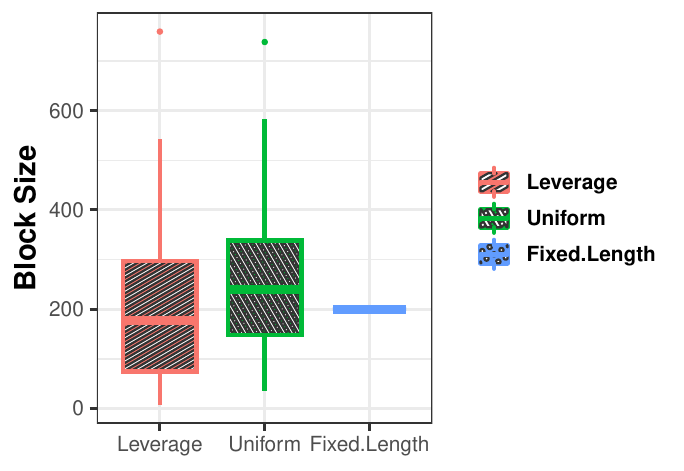}  \\
                 \scriptsize  1&\scriptsize$1.5\times 10^5$ &\includegraphics[width=3.2cm]{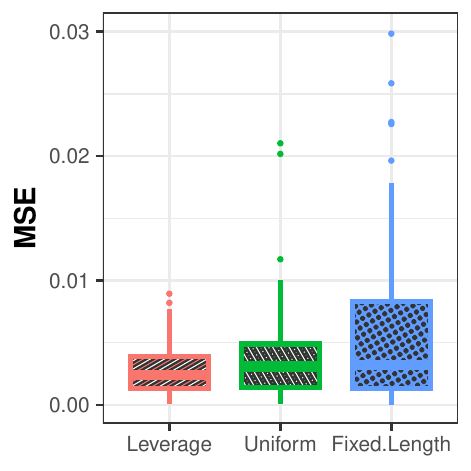}  & \includegraphics[width=3.2cm]{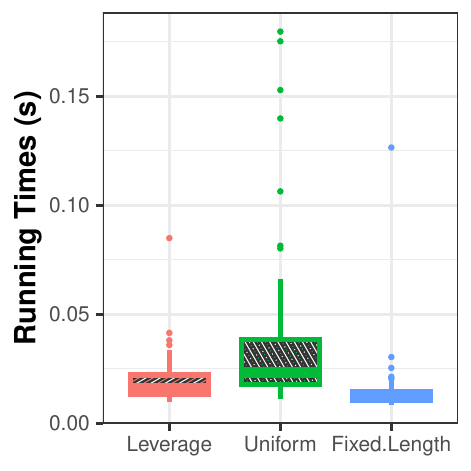}  &  \includegraphics[width=4.6cm]{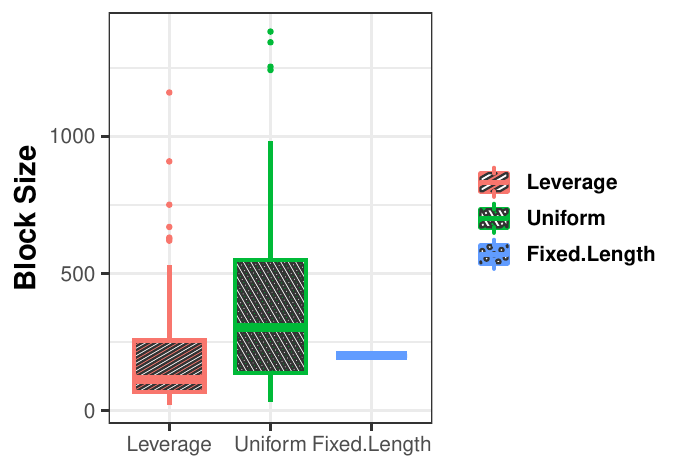}  \\
                    \scriptsize   $-1$&\scriptsize$1.5\times 10^5$ &\includegraphics[width=3.2cm]{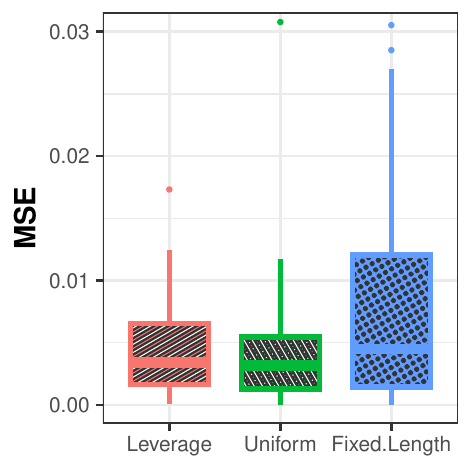} & \includegraphics[width=3.2cm]{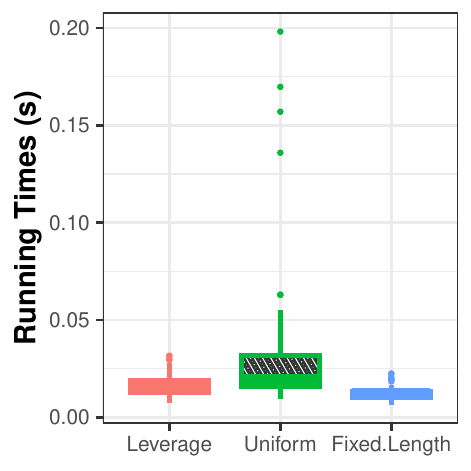}   &  \includegraphics[width=4.6cm]{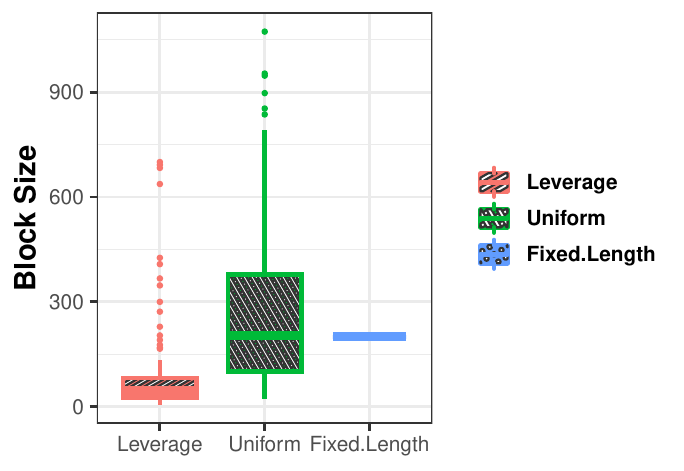}  \\
            \bottomrule
        \end{tabular}
    \end{figure}

    \underline{Model specification}. The first example that we present concerns the streaming AR($1$) series. We generate the data stream from an AR($1$) model with five different values of $\beta$ gradually changing from stable to unstable cases, i.e., $\beta = -1, -0.9, -0.5,-0.3,0.99,1$. The innovations are generated independently and identically from the $t$-distribution with $4$ degrees of freedom. We collect pilot data of size $n_0=200$. As discussed in Section~\ref{hyperparameter}, we use BIC to estimate the model order $p$.
The values of information threshold $c$ are reported in Figure~\ref{tbl:AR1_t4}.

\underline{Assessment criteria}. 
The mean squared error (MSE) of the parameter estimation is investigated, as shown in the third column of Figure~\ref{tbl:AR1_t4}.  
The total running time, which adds up training times in initialization, data selection, sequential expansion, and estimation steps, is reported in the fourth column of Figure~\ref{tbl:AR1_t4}. The block size is reported in the last column of Figure~\ref{tbl:AR1_t4}. 
All results are based on $100$ replications for each setting.


\underline{Comparison of methods}. 
Two natural benchmarks are used to compare with SLS ({Leverage}) for all cases exposed. ``Uniform Sequential Sampling'' ({Uniform}) refers to sequential sampling with starting points chosen randomly with equal probability (i.e., Bernoulli trials with equal probability), and ``Fixed Length Sampling'' ({Fixed.Length}) refers to fixed block size sampling starting at $t=n_0+1$. 
By comparing with ``Uniform Sequential Sampling'', we demonstrate the advantage of leverage-based independent Bernoulli trials for choosing the starting points as an online algorithm; while by comparing with the  ``Fixed Length Sampling'', we illustrate the the advantage of the SLS method in parameter estimation.


Figure~\ref{tbl:AR1_t4} shows that, on the parameter estimation measured by MSE, the SLS performs as well or better than other methods, especially for $|\beta|$ near $1$. Notably, the running time of SLS is significantly less than that of {Uniform} method, which reflects the computational advantage of the leverage based SLS method. The {Fixed.Length} method as a benchmark does not require any effort in selecting data points, which provides a lower bound on the running time given the similar block size.
The block size of the SLS method is also smaller than that of the {Uniform} sequential sampling, which verifies the sample efficiency result of the SLS shown in Proposition~S.2.2 of~\cite{SupplementAOAS2094} if degenerated to linear AR model. The sample efficiency of SLS explains the running time advantage since the size of the sample dominates the entire computation process. The advantages in computational efficiency (running time) and sample efficiency (block size) are more clear for unstable cases when $|\beta|$ is near 1. The {Fixed.Length} method of block size $200$ provides a reference for comparison.


\subsection{Sensitivity analysis for pilot data size $n_0$}\label{pilot_sen}
\begin{figure}[htb]
	\centering    \includegraphics[width=5.5in]{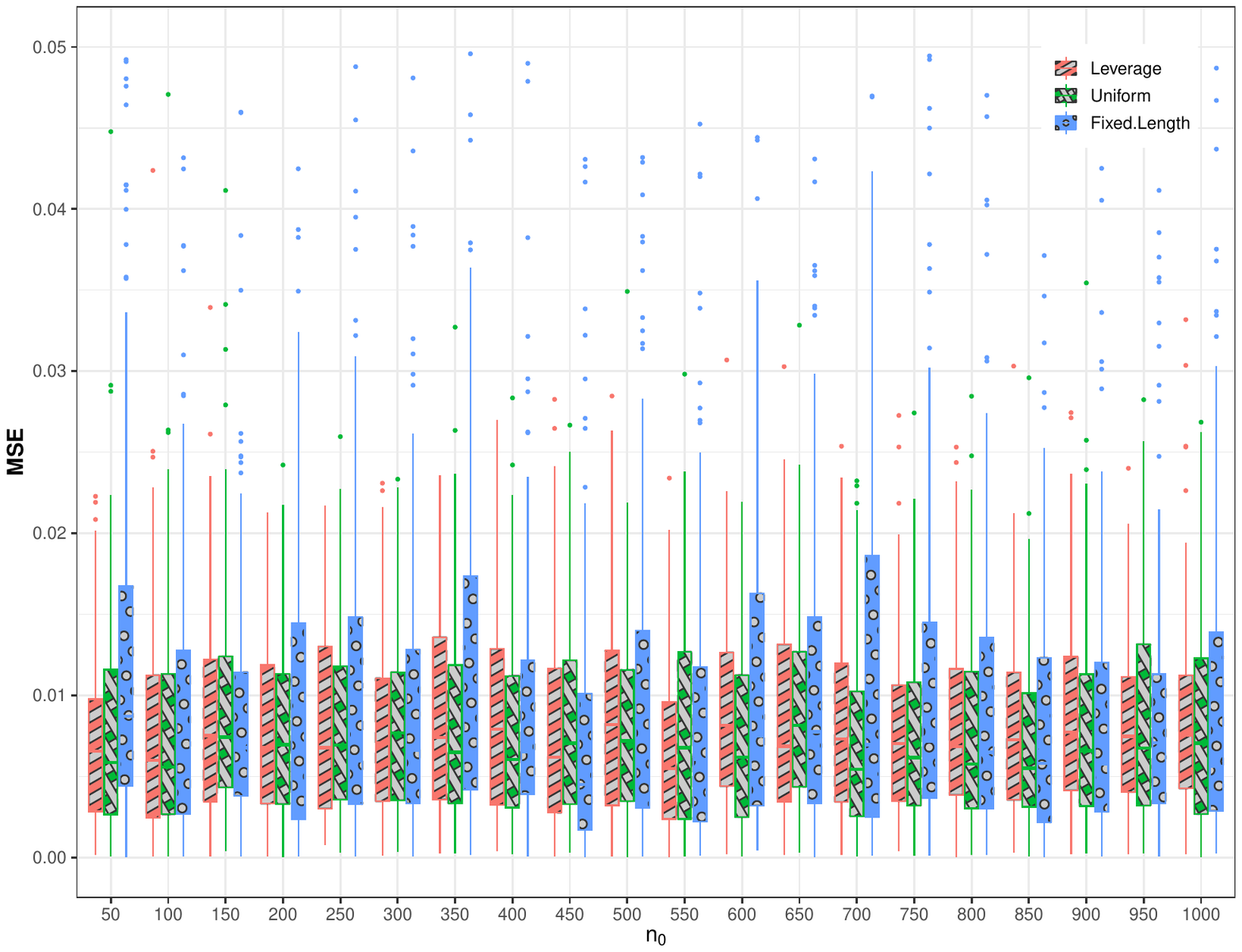}  \caption{Comparing MSE of parameter estimation for AR(1) stream on different pilot data size $n_0$ with $\beta=0.99$, observed with Sequential Leveraging Sampling and other two benchmark methods, on $100$ independent replications.
}\label{pilot_AR1}
\end{figure}

In order to understand the sensitivity and influence of the pilot data size $n_0$ on the performance of the SLS, we conduct simulation studies on comparing the parameter estimation accuracy with various values of $n_0$ for the same AR($1$) stream. In Figure~\ref{pilot_AR1}, we show the results from the unstable or nearly non-stationary AR($1$) stream with $\beta=0.99$. The results suggest that the inference performances of sampling methods are relatively insensitive to the pilot data size $n_0$.

\section{Summary}\label{diss}
In this article, we present an online data selection method called Sequential Leveraging Sampling (SLS) for streaming time series data. The SLS takes one block of consecutive time points, referred to as the SLS block, as a snapshot of the streaming data, thereby preserving the dependence structure of the selected data. The data selection algorithm involves determining the starting time and stopping time. The starting time of the SLS block is determined based on leverage scores of the streaming data, which capture the influential time points of the data stream. The stopping time is chosen according to the sequential stopping rule, which provides theoretical guarantees for estimating the properties of the time series model based on the SLS block.
Simulation and real data analysis on linear and nonlinear autoregressive models demonstrate that our SLS method is capable of efficiently processing streaming time series data in real time.

\subsection*{
Scope of applicability of SLS method}
The proposed SLS method is designed for locally stationary autoregressive time series and can be extended to independent data modeled by dynamic linear systems. While the leverage-based sampling component is easily adapted to independent settings, its theoretical properties require re-examination.

Each SLS block mimics the local data-generating process, relying on the stationarity of the block and assuming a linear, invertible autoregressive structure. This encompasses ARMA models and allows for infinite-order AR representations. The method is computationally efficient for large-scale streaming data while maintaining inferential accuracy, though some information loss is inevitable due to subsampling.

SLS assumes data recorded at regular time intervals. For high-frequency, irregularly spaced series, temporal aggregation may be needed to align frequencies and account for induced non-stationarity.

\subsection*{Future work}
The proposed SLS method demonstrates how leverage scores can guide importance sampling for dependent data, offering a more efficient and accurate alternative to simple random sampling in time series settings. Its algorithmic design and empirical performance highlight its advantage in online streaming data analysis. SLS naturally extends to multivariate time series and varying coefficient AR models~\citep{hallin1978mixed,dahlhaus1997fitting,dahlhaus1998optimal,tiao1989model}, and the leverage-based approach can be generalized to massive dependent data, such as spatial or spatiotemporal data. It also shows promise for applications in nonparametric regression, kernel learning, and matrix approximation.

\begin{acks}[Acknowledgments]
The authors would like to thank the two anonymous referees, the Associate Editor, and the Area Editor for their constructive comments that improved the quality of this paper. The authors are also grateful to Shuyang Bai, Chao Song, and Xinlian Zhang for their helpful and constructive suggestions. Additionally, the authors thank Xiaowei Chen and Fangyu Li for kindly providing the seismic data.
\end{acks}

\begin{funding}
This research was supported by the U.S.  National Science Foundation (NSF) and the U.S. National Institutes of Health (NIH). Rui Xie and Ping Ma  are supported in part by NSF grants DMS-1309665,  DMS-1438957,  DMS-1903226,  DMS-1925066, DMS-2124493, and NIH grants R01GM122080, R01GM113242. Ping Ma is in addition supported in part by NSF grants DMS-2311297, DMS-2319279, DMS-2318809 and NIH grant R01GM152814. Wei Biao Wu's research is partially supported by U.S. National Science Foundation under grant NSF DMS-2311249. T. N. Sriram 's research is partially supported by NSF grant DMS-1309665.  
\end{funding}

\begin{supplement}
\stitle{Supplement to ``Online Sequential Leveraging Sampling Method for Streaming Autoregressive Time Series with Application to Seismic Data''.} 
\sdescription{Lemmas, proofs, additional simulation studies, the extension of the SLS method to non-linear time series models, and other necessary results are included in the Supplementary Material~\cite{SupplementAOAS2094}.}
\end{supplement}


\bibliographystyle{imsart-nameyear} 
\bibliography{allref,mamahyu,TN_NSF2020seqlev,nts}       





\end{document}